\documentclass[a4paper,aps,prl,floatfix,superscriptaddress,nofootinbib,twocolumn]{revtex4-1}
\usepackage{bbm, amsmath, graphicx, mathrsfs, float}

\DeclareMathOperator*{\argmin}{arg\,min}
\newcommand{\bx}{\mathbf x}
\newcommand{\bt}{\mathbf t}

\begin{document}

\title{The stochastic matching problem}
\author{F.~Altarelli}
\affiliation{Physics Department and Center for Computational Sciences, Politecnico di Torino, Corso Duca degli Abruzzi 24, 10129 Torino, Italy}
\affiliation{Collegio Carlo Alberto, Via Real Collegio 30, 10024 Moncalieri, Italy}
\author{A.~Braunstein}
\affiliation{Human Genetics Foundation, Torino, via Nizza 52, 10126 Torino, Italy}
\affiliation{Physics Department and Center for Computational Sciences, Politecnico di Torino, Corso Duca degli Abruzzi 24, 10129 Torino, Italy}
\author{A.~Ramezanpour}
\affiliation{Physics Department and Center for Computational Sciences, Politecnico di Torino, Corso Duca degli Abruzzi 24, 10129 Torino, Italy}
\author{R.~Zecchina}
\affiliation{Physics Department and Center for Computational Sciences, Politecnico di Torino, Corso Duca degli Abruzzi 24, 10129 Torino, Italy}
\affiliation{Human Genetics Foundation, Torino, via Nizza 52, 10126 Torino, Italy}
\affiliation{Collegio Carlo Alberto, Via Real Collegio 30, 10024 Moncalieri, Italy}

\begin{abstract}
The matching problem plays a basic role in combinatorial optimization and in statistical mechanics. In its stochastic variants, optimization decisions have to be taken given only some probabilistic information about the instance.
While the deterministic case can be solved in polynomial time, stochastic variants are worst-case intractable.
We propose an efficient method to solve stochastic matching problems which combines some features of the survey propagation equations and of the cavity method. 
We test it on random bipartite graphs, for which we analyze the phase diagram and compare the results with exact bounds. Our approach is shown numerically to be effective on the full range of parameters, and to outperform state-of-the-art methods.
Finally we discuss how the method can be generalized to other problems of optimization under uncertainty.
\end{abstract}

\maketitle

One important aspect of the statistical physics approaches to disordered systems is the broad range of their interdisciplinary applications. Systems with frustration, structural disorder and uncertainties are in fact ubiquitous in many fields of science and their study has greatly benefited from the algorithms which have emerged at the interface between statistical physics of disordered systems and computer science. 

One of the key problems has been the so called matching problem \cite{matching-lovasz}, which for the case of random instances was among the first to be solved by statistical physics methods \cite{mezard-parisi-matching-1987} and later by rigorous mathematical techniques \cite{aldous-matching}.
Matching is a constituent part of many problems in different fields, ranging from physics (dimer models \cite{dimer-models}), to computer science (vision \cite{computer-vision}), economics (auctions \cite{auctions}) and computational biology (pattern matching \cite{biology}). It can be formulated simply (given a graph, find the largest possible subset of edges without common vertices), and is of polynomial complexity \cite{papadimitriou2003computational}.

The stochastic version of matching is a basic example of {\it optimization under uncertainty} \cite{prekopa1995stochastic, Birge}, which consists in finding the minimum of a cost function depending on some stochastic parameters, given just some partial information about their value. Most real-world optimization problems involve uncertainty: the precise value of some of the parameters is often unknown, either because they are measured with insufficient accuracy, or because they are stochastic in nature and determined only \emph{after} some decisions have been taken. The objective of the optimization process is thus to find solutions which are optimal in some probabilistic sense, a fact which introduces fundamental conceptual and computational challenges. 
Stochastic matching problems are in fact known to belong to higher computational complexity classes ranging from NP-hard to PSPACE-complete \cite{papadimitriou2003computational} depending on how stochasticity is introduced.

Here we apply a new method for stochastic optimization problems to the two-stage matching problem.
This new method, which builds on the formalism of Survey Propagation (SP) \cite{mezard2002random, braunstein2004survey, braunstein2005survey} and of the cavity method, is partly analytic and allows to optimize the expectation of a stochastic cost function by estimating the statistics of its minima, without resorting to explicit (and costly) sampling techniques.

In the following we define the problem, describe the method we propose for solving it, and discuss its phase diagram. We find that for large connectivity the problem enters a computationally ``hard'' phase where standard heuristics fail. In particular, we perform a detailed comparison with Stochastic Programming using state-of-the-art solvers. While our method has a good performance in both phases, Stochastic Programming turns out to be impractically slow in the region of large connectivity, and to have a significantly worse performance than our method in the region of small connectivity. Finally, we report about applications to problems which are NP-hard also in their deterministic setting.

\paragraph{The two-stage stochastic matching problem.}

We study a variant of the problem introduced in \cite{Kong, Katriel, Escoffier}, where it is shown to be NP-complete. We are given a bipartite graph $G = (L, R, E)$ with $L$ further partitioned in $L_1$ and $L_2$, and a set of independent probabilities ${\bf p} = \{p_{l_2} \in \left] 0,1 \right[, l_2 \in L_2\}$. The nodes in $L_1$ are deterministic, while the nodes in $L_2$ are stochastic: $l_2 \in L_2$ will be available for matching with probability $p_{l_2}$. In the first stage the nodes in $L_1$ are matched, knowing only the probabilities ${\bf p}$. In the second stage, the available nodes in $L_2$ are extracted according to ${\bf p}$ and they are matched. The objective is to maximise the size of the final matching.

We introduce two sets of binary variables, $\bx_1 = \{x_{l_1 r} \in \{0,1\}, (l_1 r) \in E : l_1 \in L_1\}$ and $\bx_2 = \{x_{l_2 r} \in \{0,1\}, (l_2 r) \in E : l_2 \in L_2\}$, to represent the possible $M \subset E$, with $x_{lr} = 1$ iff $(lr) \in M$. We also introduce a set of binary parameters $\bt = \{t_{l_2} \in \{0,1\}, l_2 \in L_2\}$ with $t_{l_2}=1$ iff $l_2$ is available for matching in the second stage. We define an energy function $\mathcal E(\bx_1, \bx_2, \bt)$ counting the number of unmatched vertices among the available ones. The problem consists in finding
\begin{equation}
\label{two-stage}
  \bx_1^* = \argmin_{\bx_1} \mathbbm E_{\bt} \min_{\bx_2} \mathcal E(\bx_1, \bx_2, \bt)
\end{equation}
subject to the matching constraints $\sum_{l \in \partial r} x_{l r} \leq 1 \ (\forall r \in R)$, $\sum_{r \in \partial l_1} x_{l_1 r} \leq 1 \ (\forall l_1 \in L_1)$ and $\sum_{r \in \partial l_2} x_{l_2 r} \leq t_2 \ (\forall l_2 \in L_2)$, where $\partial r = \{l \in L : (lr) \in E\}$ and similarly for $\partial l_1$ and $\partial l_2$.

The main difficulty of the problem stems from the fact that $\mathbbm E_\bt \min_{\bx_2} \mathcal E(\bx_1, \bx_2, \bt)$ has a highly non trivial dependence on $\bx_1$. In order to overcome this difficulty, we shall use the cavity method to first compute the minimum energy relative to $\bx_2$ for \emph{fixed} $\bx_1$ and $\bt$, and then to compute the average over $\bt$ of this quantity.

\paragraph{Minimizing relative to $\bx_2$ for fixed $\bx_1$ and $\bt$.}

Once $\bx_1$ is determined and the stochastic parameters $\bt$ are set, it is straightforward to find the optimal $\bx_2$. A possible way of doing this is by Max-Sum (MS), as discussed in \cite{Mezard-Zdeborova}. We introduce the cavity fields $u_{l_2 r}$ propagating from $l_2 \in L_2$ to $r \in R$ and $h_{l_2 r}$ propagating in the opposite direction. The MS equations are $u_{l_2r} = - \max[-1, \, \max_{r' \in \partial l_2 \setminus r} h_{l_2 r'}]$ and $h_{l_2 r} = -\max[-1, \, \max_{l'_2 \in \partial r \setminus l_2} u_{l_2 r}]$. These equations can be solved by iteration, and their solution allows to compute $\mathcal E^*(\bx_1, \bt) = \min_{\bx_2} \mathcal E(\bx_1, \bx_2, \bt)$, which is found to be
\begin{align}
\label{E x1 t}
  &\mathcal  E_1(\bx_1) - \sum_{l_2 \in L_2} \max[-1,\, \max_{r \in \partial l_2} h_{l_2 r}] + \\
  &- \sum_{r \in R}\max[-1,\, \max_{l_2 \in \partial r} u_{l_2 r}] + \hspace{-3mm} \sum_{(l_2 r) \in E : l_2 \in L_2} \hspace{-3mm} \max[0,\, h_{l_2 r} + u_{l_2 r}] \nonumber
\end{align}
where $\mathcal E_1(\bx_1)$ is the energy contribution of $L_1$ nodes and is constant relative to $\bx_2$.

A difficulty can arise if the solution is not unique and different solutions have different energies. In this case, only one of the solutions will correspond to the actual minimum. The following argument (confirmed by numerical investigations) suggests that this is not a problem.

The MS equations are closed for cavity fields with support in $\{-1, 1\}$, and also for cavity fields with support in $\{-1, 0, 1\}$. Solutions with other supports can exist for finite size instances and for appropriate initial conditions, but we have verified numerically that they disappear in the infinite size limit, so we shall ignore them.

Let us consider (as in \cite{Mezard-Zdeborova} for the non-bipartite case) the uniform ensemble of instances with poissonian degree distribution and average degree $c$, in the infinite size limit. The average fraction $p^u_+$ of cavity fields $u_{l_2r}$ that take the value $+1$ satisfies the equation $p^u_+ = \exp[-c \exp(-c p^u_+)]$. Also $(1 - p^u_-)$, $p^h_+$ and $(1 - p^h_-)$ must satisfy the same equation. In the case of bipartite graphs $p^h_+$ can be different from $p^u_+$ (and $p^h_-$ from $p^u_-$), and this is a notable difference relative to the non-bipartite case. In any case, $p^h_+$ and $p^h_-$ are determined from $p^u_+$ and $p^u_-$.

For $c < e$ the equation admits an unique solution, which implies that $p^u_+ = (1 - p^u_-)$ and $p^h_+ = (1 - p^h_-)$, meaning that the cavity fields have support over $\{-1, 1\}$. This unique distribution of cavity fields will correspond to an \emph{essentially} unique fixed point of MS: it is possible that some disconnected components (with finite size) admit several fixed points, but the fixed point of the $\mathrm O(N)$ component (which dominates the energy) is unique. This is confirmed by numerical simulations.

For $c > e$, the situation is more complicated: the equation $x = \exp[-c \exp(-c x)]$ admits 3 solutions, and $p^u_+$ can be different from $1 - p^u_-$ (and $p^h_+$ from $1 - p^h_-$). The condition $p^u_+ + p^u_- \leq 1$ implies that $p^u_+ \leq 1 - p^u_-$, so the total number of solutions will be at most 6. Only some of these possible solutions will correspond to positive values of the energy (and the other ones can be dismissed), and only one of them will be the correct one. We have studied in detail the case for $c = 5$: the number of solutions corresponding to positive energy is 3, and remarkably the value of the energy is the same for all of them. One of the solutions has support on $\{-1, 0, 1\}$ , corresponding to the 1-RSB case in the non bipartite case \cite{Mezard-Zdeborova},  and the remaining two have support in $\{-1, 1\}$.

On finite size instances, we have verified numerically that these 3 fixed points can always be obtained by chosing appropriate initial conditions. Their energies are close to each other, but not exactly the same, and the correct one is always the largest.

We conclude from this discussion that the energy computed from (\ref{E x1 t}) is correct for instances extracted with poissonian degree distributions with $c < e$ and approximately correct for instances with $c > e$. It must be noted, however, that the reduced instance to be solved in the second stage is not necessarily poissonian, as the probability that a node in $R$ is matched to a node in $L_1$ can be correlated to its degree. Moreover, it is possible that some small disconnected components have multiple solutions, that combined with the 3 solutions of the giant component give a larger number of fixed points, but these will always have energies that are approximately equal. We shall neglect these possible issues, comforted by our numerical results.

In the following we shall give the explicit computations for the case where the support of $\mathbf u$ and $\mathbf h$ is $\{-1,1\}$, but not for the case where the support is $\{-1, 0, 1\}$: even though we have implemented both cases, we have verified that the energies of the solutions obtained are almost exactly the same for all connectivities; however, the expressions for case $\{-1, 0, 1\}$ are much more complicated, and the running times are much longer.

\paragraph{Computing the average relative to $\bt$.}

To proceed with the computation of the average in (\ref{two-stage}), we note that the energy (\ref{E x1 t}) is a sum of local terms over $\bx_1$, $\mathbf h$ and $\mathbf u$, so that its average can be computed with a procedure similar to Survey Propagation (SP) \cite{mezard2002random, braunstein2004survey, braunstein2005survey}. This corresponds to a \emph{simplified} Belief Propagation (BP) for the variables $\bx_1$, $\mathbf h$, $\mathbf u$ and $\bt$, where $\bx_1$, $\mathbf h$ and $\mathbf u$ are subject to hard constraints implementing the matching conditions and the MS update equations, and where $\bt$ are subject to external fields that force them to take the marginal $p(\bt) = \prod_{l_2 \in L_2} \mathbbm P [t_{l_2} = 1] = \prod_{l_2 \in L_2} p_{l_2}$.

We introduce the probabilities $U_{lr} = \mathbbm P_\bt[u_{lr} = 1]$ propagating from left to right, and $H_{lr} = \mathbbm P_\bt[h_{lr} = 1]$ propagating from right to left. The SP-like equations for $U_{lr}$ and $H_{lr}$ are:
\begin{equation}
U_{lr} = p_l \prod_{r' \neq r} (1 - H_{lr'})\,\,,\,\,H_{lr} = \prod_{l' \neq l} \left( 1 - U_{l'r} \right)\label{SP_U}
\end{equation}
Equations (\ref{SP_U}) can be derived by observing that $U_{lr} = \mathbbm P \left[ t_l = 1 \right] \mathbbm P \left[ \left. -\max \left(-1, \, \max_{r' \neq r} h_{lr} \right) = 1 \right| t_l = 1 \right]$, and similarly for $H_{lr}$.

The average minimum energy is then computed by averaging (\ref{E x1 t}) over $\bt$ using $U_{lr}=\mathbbm P_\bt[u_{lr}=1]$, and $H_{lr}=\mathbbm P_\bt[h_{lr}=1]$:
\begin{eqnarray}
 \mathcal E^*(\bx_1) &=& \mathbbm E_{\bt} \mathcal E^*(\bx_1, \bt) = \sum_l p_l \left[2 \prod_r \left( 1 - H_{lr} \right) - 1 \right] + \nonumber \\
 && + \sum_r \left[ 2 \prod_l \left( 1 - U_{lr} \right) - 1 \right] + 2 \sum_{(lr)} H_{lr} U_{lr} \label{E_x1}
\end{eqnarray}
where the term $\mathcal E_1(\bx_1)$ is included and represented with the convention $U_{lr} \equiv H_{lr} \equiv x_{lr} \ \forall l \in L_1$.
For example, the contribution from a vertex $l \in L_2$ is $+1$ if the vertex is present and if all the incoming values of $h_{lr}$ are $-1$, which happens with probability $p_l \prod_r (1 - H_{lr})$; the same contribution will be $-1$ if the vertex is present and if there is at least one incoming value of $h_{lr}$ equal to $+1$, which happens with probability $p_l \left[1 - \prod_r (1 - H_{lr}) \right]$. The average of the contribution is then $p_l \left[ 2 \prod_r (1 - H_{lr}) - 1 \right]$. The average of all remaining terms is computed similarly.

Notice that the ``naive'' application of BP to the problem defined over the variables $\bx_1$, $\mathbf h$, $\mathbf u$ and $\bt$ (subject to the appropriate external fields), in which one would consider the pairs $(h_{lr}, u_{lr})$ as single joint variables, with cavity probabilities $\mathbbm P_{l \rightarrow r}[(h_{lr}, u_{lr})]$ and $\mathbbm P_{r \rightarrow l}[(h_{lr}, u_{lr})]$, would lead to the \emph{wrong} result for $c > e$. In fact, when the number of fixed points of the MS equations depends on $\bt$, the naive procedure would give to each of them a weight proportional to $p(\bt)$ while the correct weight is $p(\bt) / n_\bt$, where $n_\bt$ is the number of fixed points corresponding to a given $\bt$ and $p(\bt)$ is its probability. This is achieved with the SP procedure we introduced.

\begin{figure}
 \includegraphics{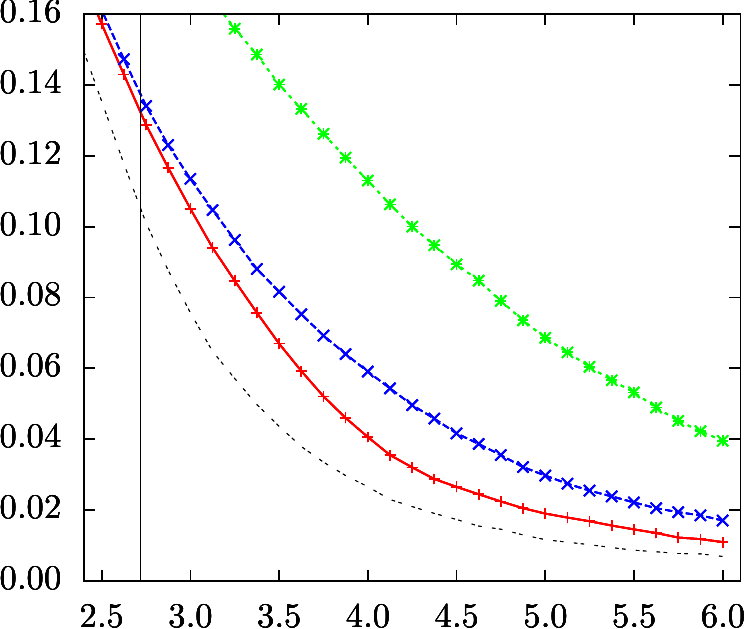}
 \caption{Average energy density vs. average connectivity of $L$ nodes. The three lines correspond (from top to bottom) to a greedy algorithm (assign $\bx_1$ as if $\bt = 0$), to a ``smart'' greedy algorithm (find the maximum-weight matching on the full instance with weights on the nodes equal to their probability to be available and assign $\bx_1$ accordingly), to the SP-derived algorithm (with $\mathbf h$ and $\mathbf u$ with support on $\{-1, 1\}$), and to the offline lower bound of the optimum (with prior knowledge of $\bt$). The vertical line is at $c = e$. Each point is an average of 50 to 100 instances, with error bars smaller than the point sizes. The instances have $|L_1| = 1\:000$ and $|L_2| = |R| = 2\:000$, with $p_l$ distributed uniformly in $]0,1]$.}
 \label{combined_plot}
\end{figure}

\paragraph{Minimizing relative to $\bx_1$.}

We can then proceed to minimize this energy, using again MS.
We consider the messages $U_{lr}$ and $H_{lr}$ as variables of a new problem and introduce the cavity messages $\mathcal U_{lr}(U) = \log \mathbbm P[U_{lr} = U]$ propagating from left to right and $\mathcal H_{lr}(H) = \log \mathbbm P[H_{lr} = H]$ from right to left. Notice that if $l\in L_1$, we will have $U_{lr}=x_{lr} \in \{0,1\}$ satisfying the matching constraints $\sum_r U_{lr} \leq 1$, while if $l \in L_2$ we will have $U_{lr} \in [0,1]$ satisfying the SP update equations (\ref{SP_U}), and similarly for $H_{lr}$. The continuous distributions over messages associated to $\bx_2$ variables can be discretized for numerical purposes.

The update equations for the messages $\mathcal U_{lr}$ and $\mathcal H_{lr}$ are obtained as usual for MS, i.e. $\mathcal U_{lr}(U_{lr}) = \max \left[ - E_{lr}(U_{lr}, H_{lr'}) + \sum_{r' \neq r} \mathcal H_{lr'}(H_{lr'}) \right]$, where $E_{lr}(U_{lr}, H_{lr'})$ is the sum of the terms in (\ref{E_x1}) containing $U_{lr}$, and where the maximisation is over the values of the incoming messages $\{H_{lr'}: r' \neq r\}$ subject to the appropriate constraints; the update of $\mathcal H_{lr} (H_{lr})$ is obtained similarly. We don't report these equations for brevity. All these maximizations can be performed efficiently by exploiting their associativity. In order to improve the convergence of the algorithm we also introduce a reinforcement term for the messages associated to edges $(l_1 r)$ with $l_1 \in L_1$ \cite{braunstein2006learning, AdWords}.

These equations can be solved by iteration starting with uniform initial conditions. These are the \emph{only} message passing equations that need to be solved numerically.
At the fixed point, the values of $\mathcal U_{lr}$ and $\mathcal H_{lr}$ provide the optimal values of $\bx_1$ by setting $x_{lr} = 1$ if and only if $[\mathcal U_{lr}(1) - \mathcal U_{lr}(0)] + [\mathcal H_{lr}(1) - \mathcal H_{lr}(0)] + 2 > 0$. Once $\bx_1$ has been assigned and the realization of $\bt$ has been extracted it is easy to perform the minimization over $\bx_2$.

\paragraph{Numerical results and comparison with other methods.}
Figure \ref{combined_plot} shows some results obtained with the SP-derived algorithm in the case where $\mathbf h$ and $\mathbf u$ have support on $\{-1, 1\}$. The case where $\mathbf h$ and $\mathbf u$ have support on $\{-1, 0, 1\}$ gives results that are very close to these.

\begin{figure}
\includegraphics[width=1.0\columnwidth]{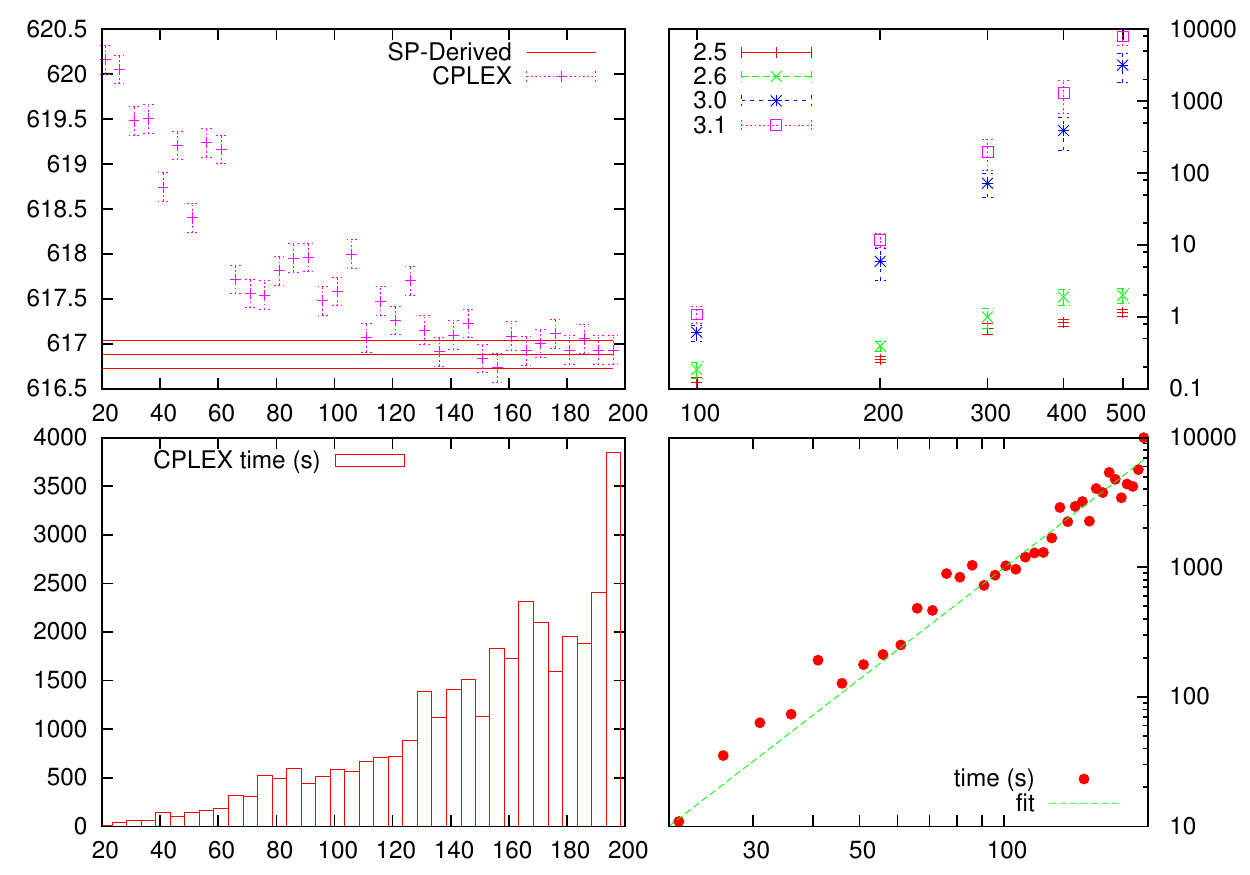}
\caption{Each point represents an instance with $|L_2|=|R|=2|L_1|$ and $p_l$ distributed uniformly in $]0,1]$. Top Left: Energy vs. number of samples $\rho$ obtained by CPLEX for $c=2.5$ and $|L_1|=1\:000$, compared with the one computed by the SP-derived algorithm. The energies and error bars were computed by resampling over $10\:000$ samples. Bottom Left: CPLEX time (seconds) vs. number of samples in the same instances. Top Right: CPLEX time as a function of $|L_1|$ for several values of $c$ and $\rho=10$. Bottom Right: Best fit of CPLEX times in Bottom Left with $f(x)=b x^a$ gives $a\simeq2.35$.\label{fig:cplex}}
\end{figure}

In order to give quantitative evidence of the potentialities of our approach for real world problems, we have made two final studies: on the one hand we have compared the performance with state-of-the art method, and on the other we have applied the method to problems which are NP-hard even in the deterministic setting. In both cases, the results perfectly corroborate our expectations.
{\bf (i)} We compared the SP-derived algorithm with two other standard approaches. The first is a greedy strategy solving a weighted matching based on $p(\mathbf t)$: even though it is very fast, its solutions are much worse (Figure \ref{combined_plot}). The second is called stochastic programming. It consists in extracting $\rho$ realizations $\bt^{1},\dots,\bt^{\rho}\sim p(\bt)$  and then solving $\min_{\bx_1} \sum_{i=1}^{\rho} \min_{\bx^{i}_2} \mathcal E(\bx_1, \bx^{i}_2, \bt^{i}) = \min_{\bx_1, \bx^{1}_2, \dots, \bx^{\rho}_2} \sum_{i=1}^{\rho} \mathcal E(\bx_1, \bx^{i}_2, \bt^{i})$ using OR techniques like linear relaxations complemented with branch-and-bound. Note that this minimization problem is NP-Complete\cite{Kong}. We employed two well known tools for this task: iLog CPLEX, a commercial, industrial strenght linear/integer programming software from IBM, and \texttt{lp\_solve}, an open source alternative. Although qualitatively similar, results with \texttt{lp\_solve} were uniformly worse than the ones of CPLEX, so we will not report them.
We observe that the results depend strongly on $\rho$ and on the average degree $c$. As expected, for fixed $c$ the quality of the solution improves as $\rho$ increases, but the running time becomes larger. For $c$ up to around $2.5$, CPLEX seems to be able to solve the problem in polynomial time in both $\rho$ and $N$, but either it is much slower than the SP-derived algorithm or it gives a significantly higher energy (depending on $\rho$). For $c=3.5$ and above, the time scaling of CPLEX worsens significantly: for $\rho = 10$, the running time increases dramatically with $|L_1|$, and for $|L_1|=1000$ CPLEX was not able to attain an optimum under a cutoff of 24 hours even for $\rho=2$. Note that the SP-derived algorithm employs around one minute.
{\bf (ii)} The method was also successfully applied to a stochastic version of the maximum weight independent set problem \cite{Sanghavi2009}, when the node's contribution to the total weight is uncertain
but its distribution is known. This is a relevant problem in communication networks with some interference constraints \cite{Shah2008}. Details will be given in \cite{long-paper}.
\begin{acknowledgments}
RZ acknowledges the EU grant n. 265496.
\end{acknowledgments}


\end{document}